# Temporal Parity-Time Symmetry for Extreme Energy Transformations

Huanan Li[1,*], Shixiong Yin[1,2,*], Emanuele Galiffi[3], and Andrea Alù[1,2,4†]

[1]*Photonics Initiative, Advanced Science Research Center, City University of New York, New York, New York 10031, USA*

[2]*Department of Electrical Engineering, City College of The City University of New York, New York 10031, USA*

[3]*The Blackett Laboratory, Imperial College London, London SW7 2AZ, UK*

[4]*Physics Program, Graduate Center, City University of New York, New York, New York 10016, USA*

*Temporal interfaces introduced by abrupt switching of the constitutive parameters of unbounded media enable unusual wave phenomena. So far, their explorations have been mostly limited to lossless media. Yet, non-Hermitian phenomena leveraging material loss and gain, and their balanced combination in parity-time (PT)-symmetric systems, have been opening new vistas in photonics. Here, we unveil the role that temporal interfaces offer in non-Hermitian physics, introducing the dual of PT symmetry for temporal boundaries. Our findings reveal unexplored interference mechanisms enabling extreme energy manipulation, and open new scenarios for time-switched metamaterials, connecting them with the broad opportunities offered by non-Hermitian phenomena.*

Time-varying media exhibit great potential for extreme wave manipulation and for next-generation technologies, drawing in recent years significant attention from the engineering and physics communities. These systems exploit temporal degrees of freedom to build higher-dimensional metamaterials [1], and enable intriguing opportunities such as Floquet topological insulators for both classical and quantum waves [2]-[10], modulation-induced non-reciprocity [11]-[17], multifunctional metasurfaces [18], overcoming several limitations of static media [19]-[27]. As an important subclass, abrupt changes in time of material properties produce temporal boundaries at which a wave can experience reflection and refraction, enabling a dual phenomenon to a spatial interface [28]-[31]. For the past few years, the effects of temporal boundaries and their application opportunities have been intensively investigated [32]-[39]. Nevertheless, the vast majority of these studies have been limited to lossless media, in which their permittivity or permeability changes in time.

---

[*] These authors contributed equally.
[†] Corresponding author: aalu@gc.cuny.edu

A parallel line of research holding significant promise in the context of exotic wave-matter interactions involves non-Hermitian systems, which capitalize on the interplay between gain and loss to enable a wealth of opportunities to control and transmit waves, realizing unconventional phenomena, most notably when gain and loss are balanced and the system obeys parity-time (PT) symmetry [40]. Unidirectional invisibility [41]-[43], laser-absorber pairs [44]-[45] and chiral state transfer [46]-[50] leverage PT symmetry and non-Hermitian singularities, leading to exotic phenomena that have no counterparts in Hermitian systems. Here, we unveil the untapped opportunities enabled by non-Hermitian temporal slabs, in which the material conductivity is abruptly changed in time. In this quest, we forge a link between the field of non-Hermitian systems and the one of time-metamaterials, enabling novel interference phenomena for extreme energy manipulation, based on the interaction of non-orthogonal counter-propagating waves for which the carried power is decoupled from the total stored energy. Of particular interest is the scenario in which paired temporal slabs obey temporal parity-time (TPT) symmetry, exhibiting a phenomenon dual to laser-absorber pairs [44]-[45] for which waves can be largely amplified or attenuated over a broad dynamical range of output power levels.

*Non-Hermitian temporal slabs* — We begin by considering wave propagation in a uniform and isotropic non-magnetic medium characterized by a real-valued (dispersion-less) relative permittivity $\varepsilon$ and electrical conductivity $\sigma$. Without loss of generality, we assume the presence of two counter-propagating waves in the $z$-direction, sharing the same wavenumber $k_z > 0$, so that the total transverse electric and magnetic field components are

$$E_x(z,t) = e^{-jk_z z}[E^+(t) + E^-(t)] + c.c.;$$
$$\eta_0 H_y(z,t) = e^{-jk_z z}[nE^+(t) - n^* E^-(t)] + c.c.,$$
(1)

with $E^+(t) = E_+ e^{j\omega t}$ and $E^-(t) = E_-^* e^{-j\omega^* t}$ being the time-dependent amplitudes of the forward and backward wave at the complex (angular) frequency $\omega = k_z c_0/n$, and $c_0 = 1/\sqrt{\mu_0 \varepsilon_0}$, $\eta_0 = \sqrt{\mu_0/\varepsilon_0}$ being the speed of light and characteristic impedance in free space, $\mu_0$ ($\varepsilon_0$) being its permeability (permittivity). The complex refractive index $n \equiv \sqrt{\varepsilon + \sigma/(j\omega \varepsilon_0)} = n' - jn''$ with its real (imaginary) part $n' = \sqrt{\varepsilon - \hat{\sigma}^2}$ ($n'' = \hat{\sigma}$) where $\hat{\sigma} \equiv \sigma/(2\varepsilon_0 k_z c_0)$, and $c.c.$ stands for complex conjugate. After averaging in space over one spatial period $2\pi/k_z$, the energy density $w_{em}(t)$ at time $t$ can be written as [51]

$$w_{em}(t) = U_i(t) + U_c(t),$$
(2)

where $U_i(t) \equiv 2\varepsilon_0 \varepsilon [|E^+(t)|^2 + |E^-(t)|^2]$ is the average energy stored in the two individual waves, and $U_c(t) \equiv 2\varepsilon_0 \text{Re}[C(t)]$ with $C(t) = n(n^* - n)E^+(t)E^-(t)^*$ represents the contribution to stored energy stemming from the interference between the two waves. $U_i(t)$ is directly proportional to the (spatially averaged) total power flow carried by the two waves $P_{tot}(t) = P^+(t) + P^-(t)$ with $P^\pm(t) = (2n'/\eta_0)|E^\pm(t)|^2$ [52], i.e., $P_{tot}(t) = U_i(t) c_0 n'/\varepsilon$, while $U_c(t)$ emerges only in non-Hermitian media ($\sigma \neq 0$) where the two counter-propagating waves are non-orthogonal. In non-Hermitian media the interference between counter-propagating waves with real-wavenumber and complex frequency is different from the most commonly studied scenario of interference of real-frequency waves, which has been explored in connection with Anderson localization [53]-[61], and for which the time-average power flow of the individual waves is not physically meaningful [62].

Non-Hermitian time boundaries arise when the conductivity is switched in time, offering unexplored avenues to control these quantities. Upon switching of the relative permittivity $\varepsilon$ and conductivity $\sigma$ of the unbounded medium, momentum $k_z$ instead of frequency is conserved. In addition, at the temporal boundary at time $t_s$ for which the material properties transition abruptly from $(\varepsilon_1, \sigma_1)$ to $(\varepsilon_2, \sigma_2)$, or equivalently from $n_1$ to $n_2$ for the complex refractive index $n$, the time-dependent amplitudes $E^\pm(t)$ of forward and backward waves vary as

$$\psi_E(t_s^+) = J_{2,1}\psi_E(t_s^-), \qquad J_{2,1} = \begin{pmatrix} \tau_{2,1} & \rho_{2,1}^* \\ \rho_{2,1} & \tau_{2,1}^* \end{pmatrix}, \tag{3}$$

where $\psi_E(t) \equiv (E^+(t), \ E^-(t))^T$ (the superscript $T$ indicates the transpose), $J_{2,1}$ is the matching matrix ensuring that the displacement field $D_x = \varepsilon E_x$ and magnetic field $H_y$ are conserved. The temporal transmission and reflection coefficients are $\tau_{2,1} = [n_1(n_1^* + n_2)]/[n_2(n_2 + n_2^*)]$ and $\rho_{2,1} = [n_1(n_1^* - n_2^*)]/[n_2^*(n_2 + n_2^*)]$. In the limiting case $\sigma_1 = \sigma_2 = 0$, these coefficients revert to well-known results for Hermitian time boundaries [28]-[29],[63]-[64]. Due to wave interference, a non-Hermitian temporal boundary can produce intriguing scattering events, including the nontrivial decoupling of total power flow and total stored energy. Consider first the special scenario when $\varepsilon_2 = \varepsilon_1$ and $\sigma_2 \neq \sigma_1 = 0$, i.e., the time boundary involves only an abrupt change in conductivity. The temporal transmittance $T^+$ and reflectance $T^-$ are $T^\pm \equiv P^\pm(t_s^+)/P^+(t_s^-) =$ $\left(1 \pm \sqrt{1 - \hat{\sigma}_2^2/\varepsilon_1}\right) / \left(2\sqrt{1 - \hat{\sigma}_2^2/\varepsilon_1}\right)$, and thus the total power flow $T^+ + T^- = \frac{1}{\sqrt{1-\hat{\sigma}_2^2/\varepsilon_1}} \neq 1$ changes after the switching, even though the total stored energy in the waves does not, $w_{em}(t_s^+) = w_{em}(t_s^-)$, since the permittivity is continuous across the temporal boundary. As shown later, this counterintuitive result can be explained by the exotic wave interference emerging in non-Hermitian temporal slabs and, when combined with TPT symmetry, it enables extreme manipulation of the total power flow. We define TPT symmetry here as the dual in time domain of spatial PT-symmetry, in which we replace the conventional parity and time reversal operators with their temporal analogues $\mathcal{P}_t$ and $\mathcal{T}_t$. The parity $\mathcal{P}_t$ operation is defined as a reversal along the time axis $t \to -t$, while the $\mathcal{T}_t$ operation sets $z \to -z$ and it performs complex conjugation via the operator $T$.

***Temporal scattering formalism and TPT-symmetric slabs*** — Before analyzing TPT-symmetric bilayers, we need to introduce a temporal scattering formalism accounting for the evolution of $\psi_E(t)$ in non-Hermitian temporal slabs. Using Eq. (1), $\psi_E(t)$ in medium $l$ evolves from time $t_1$ to $t_2 = t_1 + \Delta t_{21}$ as

$$\psi_E(t_2) = F_l(\Delta t_{21})\psi_E(t_1), \qquad F_l(\Delta t_{21}) = \begin{pmatrix} e^{j\omega_l \Delta t_{21}} & 0 \\ 0 & e^{-j\omega_l^* \Delta t_{21}} \end{pmatrix}, \qquad (4)$$

where $F_l(\Delta t_{21})$ is the propagation matrix and the frequency $\omega_l = k_z c_0 / n_l$ generally assumes complex values. For non-Hermitian temporal slabs involving multiple switching events, the matching matrix $J_{2,1}$ in Eq. (3) and the propagation matrix $F_l$ in Eq. (4) constitute the building blocks of the temporal transfer matrix $M$ connecting the state $\psi_E(t)$ at two different times. We consider the general scenario in Fig. **1**(a), based on which an unbounded medium switches abruptly at time $t_i = 0, \Delta t$ and $t_f = 2\Delta t$ from medium $l$ to $l + 1$, with $l = 1, 2, 3$. We assume that media 1 and 4 are Hermitian ($\sigma_1 = \sigma_4 = 0$) with same permittivity $\varepsilon_1 = \varepsilon_4 > 0$, while media 2 and 3 have equal relative permittivity $\varepsilon_2 = \varepsilon_3 > 0$ but balanced loss and gain, i.e., $\sigma_2 = -\sigma_3 > 0$, and thus TPT-symmetric. On average, we expect these media to compensate each other in terms of energy decay or amplification, in a way dual to PT-symmetric slabs in their symmetric phase [41]-[42]. Following Eqs. (3) and (4), the total transfer matrix $M_{tot}$ in this geometry, defined via $\psi_E(t_f^+) = M_{tot}\psi_E(t_i^-)$, reads

$$M_{tot} = J_{4,3}F_3(\Delta t)J_{3,2}F_2(\Delta t)J_{2,1}, \tag{5}$$

which consists of the propagation matrices $F_l(\Delta t)$ in media $l = 2, 3$, and the matching matrices $J_{l+1,l}$ between different media, and it determines the full evolution of the counter-propagating plane waves right before the first switching event at $t_i = 0$.

Alternatively, the scenario of Fig. **1**(a) can be described by the *temporal* scattering matrix $S = \begin{pmatrix} r_L & t_R \\ t_L & r_R \end{pmatrix}$, which obeys the relation $|s^-(t_f^+)\rangle = S|s^+(t_i^-)\rangle$ with the input $|s^+(t_i^-)\rangle = \psi_E(t_i^-)$ and output $|s^-(t_f^+)\rangle = P\psi_E(t_f^+)$, $P \equiv \begin{pmatrix} 0 & 1 \\ 1 & 0 \end{pmatrix}$, and it is related to $M_{tot}$ via the relation $S = PM_{tot}$. Interestingly, we find that the dynamics of non-Hermitian temporal slabs are always invariant under the $\mathcal{T}_t$ operation, leading to the universal constraint $PTSPT = S$, or equivalently $r_R = r_L^*$ and $t_R = t_L^*$ [65]. It follows that the dynamics of any TPT-symmetric structure, defined as above, are necessarily invariant under both $\mathcal{P}_t$ and combined $\mathcal{P}_t\mathcal{T}_t$ operations, and the corresponding temporal $S$ matrix obeys the fundamental relationship $PTSPT = S^{-1}$, i.e., the time analogue of conventional PT-symmetric scattering systems. Nevertheless, due to the constraint imposed by the symmetry under the $\mathcal{T}_t$ operation, we parametrize the $S$ matrix of TPT-symmetric structures as

$$S = \begin{pmatrix} jb & a^* \\ a & -jb \end{pmatrix}, |a|^2 - b^2 = 1, b \in \mathcal{R}, a \in \mathcal{C}. \tag{6}$$

This result ensures that the eigenvalues are $\pm 1$: different from conventional PT-symmetry, TPT-symmetric structures are always in their symmetric phase, and the temporal $S$ matrix in Eq. (6) and the $PT$ operator share the same set of eigenvectors. Eq. (6) also implies that TPT-symmetric structures cannot attenuate a single input wave, since the temporal transmittance $T^+ = |a|^2 \geq 1$.

***Extreme energy transformations in TPT-symmetric temporal bilayers*** — Despite the fact that TPT-symmetric systems are always in their symmetric phase with unimodular eigenvalues, highly nontrivial energy manipulation is possible exploiting wave non-orthogonality. To demonstrate these opportunities, we consider equal-intensity counter-propagating input waves with a variable relative phase $\phi$ at time $t_i^- = 0^-$ [see Fig. **1**(a)], i.e., $|s^+(t_i^-)\rangle = (1/\sqrt{2}, \; e^{j\phi}/\sqrt{2})^T$, and investigate the normalized total output power flow $\hat{P}_{tot} \equiv P_{tot}(t_f^+)/P_{tot}(t_i^-)$ at time $t_f^+ = (2\Delta t)^+$. In our scenario, $\hat{P}_{tot}$ depends on four dimensionless real-valued parameters: the relative phase $\phi$, the (normalized) temporal slab duration $\Delta_f \equiv 2\Delta t \omega_1$, and the real and imaginary parts $\hat{n}_2'$ and $\hat{n}_2''$ of the refractive index ratio $\hat{n}_2 \equiv n_2/n_1 = \hat{n}_2' - j\hat{n}_2''$. In addition, $\hat{P}_{tot}$ is a periodic function of $\phi$ and $\Delta_f$ for fixed $\hat{n}_2$, i.e., $\hat{P}_{tot}(\phi + 2\pi, \Delta_f) = \hat{P}_{tot}(\phi, \Delta_f + 2\pi|\hat{n}_2|^2/\hat{n}_2') = \hat{P}_{tot}(\phi, \Delta_f)$.

In analogy to laser-absorber pairs in PT-symmetric systems, we explore the global minimum $\hat{P}_{tot}^{min}$ and maximum $\hat{P}_{tot}^{max}$ of the total normalized output power $\hat{P}_{tot}(\phi, \Delta_f)$ as we vary the relative phase between the input waves $\phi$ and the temporal slab thickness $\Delta_f$. Remarkably, a critical-point analysis of $\hat{P}_{tot}$ (see below) indicates that the global minimum $\hat{P}_{tot}^{min}$ and maximum $\hat{P}_{tot}^{max}$ are achieved for the same value $\Delta_f^{min} = \Delta_f^{max}$. Since $\hat{P}_{tot}(\phi, \Delta_f)$ is a continuous function, the maximum dynamical range between $\hat{P}_{tot}^{max}$ and $\hat{P}_{tot}^{min}$ can be reached for this value of $\Delta_f$ by modulating the relative phase $\phi$. Next we study the variation of these extreme values as the imaginary part $\hat{n}_2''$ changes.

Assuming a fixed value of $\hat{n}'_2 > 1$, a perturbation analysis with $\hat{n}''_2 \to 0^+$ sheds light on the extreme values that $\hat{P}_{tot}$ can attain. When $\hat{n}''_2 = 0$, the TPT-symmetric temporal bilayer in Fig. **1**(a) is a uniform Hermitian temporal slab, and $\hat{P}_{tot}$ changes at times $t_i = 0$ and $t_f = 2\Delta t$, together with the stored energy in the system. Owing to the equal-intensity inputs $|s^+(t_i^-)\rangle$ and the symmetry of matching and propagation matrices in Eqs. (3) and (4), the intensities of the counter-propagating waves within each slab are equal. This observation together with Eq. (3) allows us to obtain the optimal relative phase 0 and $\pi$ (or $\pi$ and 0) of the two waves at time $t_i^-$ and $t_f^-$ respectively in order to achieve maximum (or minimum) energy extraction from the waves, yielding $\hat{P}_{tot}^{min} = 1/\hat{n}'^2_2$ for $\phi_{min} = 0$ and $\Delta_f^{min} = \hat{n}'_2 \pi/2$ for the global minimum, and $\hat{P}_{tot}^{max} = \hat{n}'^2_2$ for $\phi_{max} = \pi$ and $\Delta_f^{max} = \Delta_f^{min}$ for the global maximum. For TPT-symmetric slabs, we can now write a perturbation series in $\hat{n}''_2$ for $\Delta_f^{max} = \Delta_f^{min} = \hat{n}'_2\pi/2 + \hat{n}''_2(\hat{n}'^2_2 + 1)/(\hat{n}'^2_2 - 1) + \hat{n}''^2_2 \pi/(2\hat{n}'_2) + O(\hat{n}''^3_2)$ and $\phi_{max} - \pi = \phi_{min} = 2\hat{n}''_2/(\hat{n}'^2_2 - 1) + O(\hat{n}''^3_2)$, yielding the global extreme values

$$\frac{1}{\hat{P}_{tot}^{max}} = \hat{P}_{tot}^{min} = \frac{1}{\hat{n}'^2_2} - \frac{2\hat{n}''_2}{\hat{n}'^3_2} + \frac{2\hat{n}''^2_2}{\hat{n}'^4_2(1 - \hat{n}'^2_2)} + O(\hat{n}''^3_2). \tag{7}$$

In Fig. **1**(b), the extreme values of $\hat{P}_{tot}$ when $\hat{n}'_2 = 2$, i.e., $\hat{P}_{tot}^{min}$ (blue down-pointing triangles) and $\hat{P}_{tot}^{max}$ (red up-pointing triangles) in Eq. (7), versus $\hat{n}''_2 \in (0, 0.4)$ are plotted against rigorous numerical calculations (blue and red solid lines), showing good agreement. The non-Hermitian nature of the TPT-symmetric bilayer drastically enhances the dynamical range of achievable minimum and maximum power after the switching events. To explain this phenomenon, we evaluate the time-dependent evolution of the (normalized) stored energy density $\hat{w}_{em}(t) \equiv w_{em}(t)/w_{em}(0^-)$ and its two components $\hat{U}_i(t) \equiv U_i(t)/w_{em}(0^-)$ and $\hat{U}_c(t) \equiv U_c(t)/w_{em}(0^-)$, plotting them as a function of time for $\hat{n}_2' = 2$ and $\hat{n}_2'' = 0.2$ in Fig. 2(a) for the configuration providing $\hat{P}_{tot}^{min}$, and in Fig. 2(b) for $\hat{P}_{tot}^{max}$. The stored energy density $\hat{w}_{em}(t)$ (solid lines) is discontinuous at time $t_i$ and $t_f$, since we pump or subtract energy from the material as we switch its permittivity $\varepsilon = |n|^2$, but it is continuous in the middle switching event because here we only switch the conductivity. $\hat{U}_i(t)$ (dashed lines) and $\hat{U}_c(t)$ (empty circles) are discontinuous in the middle switching event, compensating each other. The TPT bilayer largely absorbs [Fig. 2(a)] or amplifies [Fig. 2(b)] the input waves by simply changing their relative phase, leading to a largely manipulated total power flow with $\hat{P}_{tot} = \hat{w}_{em}(t_f^+)$.

We now explore the extreme scenario in which we only switch the conductivity, such that, as we change $\hat{n}_2''$, we also vary $\hat{n}_2' = \sqrt{1 - \hat{n}_2''^2} < 1$ correspondingly. This scenario is particularly interesting since we cannot expect any variation in stored energy across any of the switching events. Yet, as anticipated above, it is actually possible to largely modify the power flow even in this regime, and the extreme values of $\hat{P}_{tot}$ read

$$\hat{P}_{tot}^{min} = \frac{(\hat{n}_2'' - 1)^2}{(\hat{n}_2'' + 1)^2}, \hat{P}_{tot}^{max} = \frac{1}{\hat{P}_{tot}^{min}} \qquad (8)$$

reached for $\phi_{min} = \pi/2$, $\Delta_f^{min} = \pi/\sqrt{1 - \hat{n}_2''^2}$ and $\phi_{max} = 3\pi/2$, $\Delta_f^{max} = \Delta_f^{min}$. Interestingly, the optimal phases do not depend on $\hat{n}_2''$ in this case. In Fig. 1(b), blue and red dashed lines show the range of available $\hat{P}_{tot}$ at its extreme values as a function of $\hat{n}_2''$. As expected, they both start from unity for small $\hat{n}_2''$, but as $\hat{n}_2''$ approaches 1, they converge to 0 and ∞ dual to a laser-absorber pair in PT-symmetric systems [see Eq. (8)], despite the fact that *all switching events preserve the stored energy*, and that the TPT-symmetric bilayers operate in their symmetric phase. The drastic variations of $\hat{P}_{tot}$ emerge here as the waves become non-orthogonal in non-Hermitian temporal slabs. In Fig. **3**, we show the two extreme scenarios for $\hat{n}_2'' = 0.2$ and $\hat{n}_2' = \sqrt{1 - \hat{n}_2''^2} \approx 0.98$, studying the evolution of normalized energies as in Fig. **2**. As expected, the energy density $\hat{w}_{em}(t)$ is continuous across every switching event, but $\hat{U}_i(t)$ and $P_{tot}(t)$ are largely modified. For $\hat{P}_{tot}^{min}$, both temporal slabs start from a maximum $\hat{U}_c(t)$, thanks to the proper phase of the incident waves, and end at the point of its maximum decay, minimizing the overall residual energy in the waves and corresponding power flows [Fig. **3**(a)]. In the dual scenario of $\hat{P}_{tot}^{max}$ [Fig. **3**(b)], on the contrary, $\hat{U}_c(t)$ in each temporal slab starts at a minimum and grows, maximizing the overall power flow at the output.

For both Figs. **2** and **3**, we performed COMSOL simulations to validate our analysis. As shown in each inset, two counter-propagating pulses with equal intensity travel against each other in a homogeneous Hermitian medium, meet and overlap at time $t = 0$ with a prescribed relative phase, experience three switching events, following the color code in the figure, and separate once they are again in the Hermitian medium. The simulations match well our model, which ignores the finite spatial extent of the wave trains. The inset of each subfigure shows time snapshots of the total $E_x(z, t)$ distribution of the counter-propagating waves at times $t = -16\pi/\omega_1, \Delta t/2, 3\Delta t/2$ and $16\pi/\omega_1$, and movies of the wave evolution are provided in [66].

*Conclusions* — In this Letter, we introduced a rigorous temporal scattering formalism for non-Hermitian temporal slabs, highlighting unexplored opportunities enabled by non-Hermiticity in time metamaterials based on nontrivial wave interference. Based on this formulation, we have introduced the analogue of PT-symmetry for temporal slabs, showcasing highly nontrivial energy manipulation features. TPT-symmetric systems are always in the symmetric phase, yet they support the dual phenomenon of laser-absorber pairs, enabling an extreme dynamical range of output power levels tuning the relative phase between input counter-propagating waves. These results may be further enhanced by exploring these wave interference phenomena in non-instantaneous materials [67] and/or in open resonators with large quality factors, providing new strategies for the design of time-metamaterial devices for extreme wave manipulation. By generalizing our approach to incorporate also spatial boundaries, it may be possible to enable other PT-symmetric phenomena, like phase transitions, in combination with the interference-driven phenomena unveiled here.

*Acknowledgements* ─ This work was supported by the Simons Foundation and the Air Force Office of Scientific Research MURI program. E.G. acknowledges support from the Engineering and Physical Sciences Research Council through an EPSRC Doctoral Prize Fellowship (grant EP/T51780X/1).

**Figures**

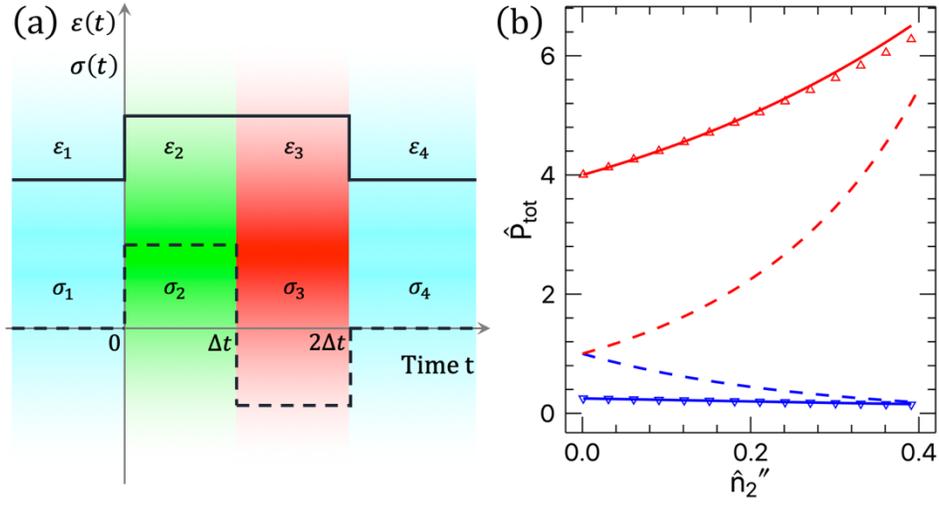

**Fig. 1.** (a) Schematic of TPT-symmetric temporal slabs. (b) $\hat{P}_{tot}^{min}$ [solid (or dashed) blue line] and $\hat{P}_{tot}^{max}$ [solid (or dashed) red line] in the case of fixed $\hat{n}_2' = 2$ (or varying $\hat{n}_2' = \sqrt{1 - \hat{n}_2''^2}$) as a function of $\hat{n}_2''$. Blue down-pointing (red up-pointing) triangles represent the perturbative result for $\hat{P}_{tot}^{min}$ ($\hat{P}_{tot}^{max}$) in Eq. (7).

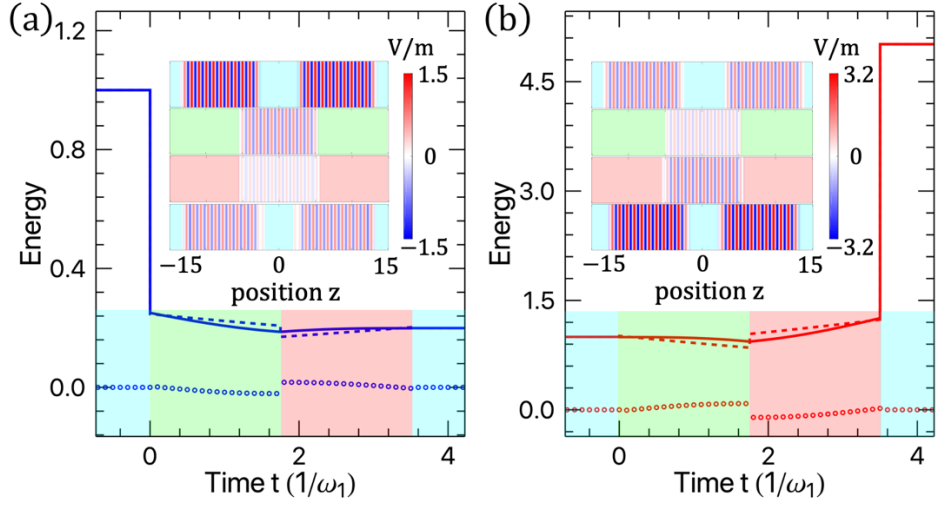

**Fig. 2.** Normalized energy $\hat{w}_{em}(t)$ (solid lines), $\hat{U}_i(t)$ (dashed lines) and $\hat{U}_c(t)$ (dotted lines) as a function of time when (a) $\phi_{min} = 0.13$, $\Delta_f^{min} = 3.5$ and (b) $\phi_{max} = 3.27$, $\Delta_f^{max} = \Delta_f^{min}$ for the extreme values of $\hat{P}_{tot}$ [see Fig. **1**(b)] in the case of $\hat{n}_2' = 2$ and $\hat{n}_2'' = 0.2$. Insets: snapshots obtained with COMSOL of $E_x(z,t)$ versus position $z$ (in units of $2\pi/k_z$) at time $t = -16\pi/\omega_1$ (1st row), $\Delta t/2$ (2nd row), $3\Delta t/2$ (3rd row) and $16\pi/\omega_1$ (4th row) for finite pulses traveling in the medium. Other free parameters are $n_1 = 1$ and $\omega_1/(2\pi) = 1$ GHz, and the colored backgrounds indicate the material properties over time. The amplitude of the two input waves is $\sqrt{2}$ V/m.

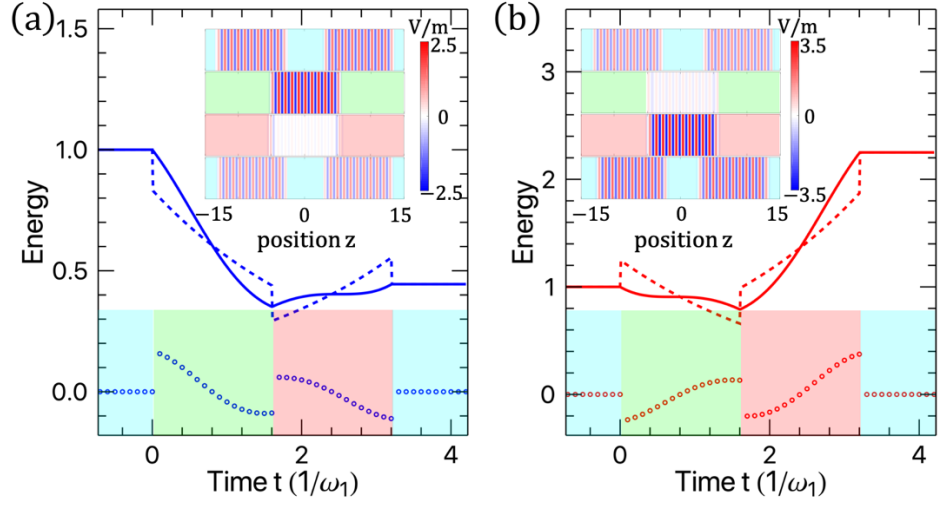

**Fig. 3.** Same as Fig. **2** for the minimum (a) and the maximum (b) of $\hat{P}_{tot}$ when $\hat{n}_2'' = 0.2$ and $\hat{n}_2' \approx 0.98$, such that we switch only the material conductivity.